# Bot Development for Social Engineering Attacks on Twitter

Jefferson Viana Fonseca Abreu[1], Jorge Henrique Cabral Fernandes[1],
João José Costa Gondim[1], Célia Ghedini Ralha

[1]Departamento de Ciência da Computação – Universidade de Brasília
Campus Universitário Darcy Ribeiro - Brasília, DF, Brazil

jeffvfa@hotmail.com, jhcf@unb.br, gondim@unb.br, ghedini@unb.br

***Abstract.*** *A series of bots performing simulated social engineering attacks using phishing in the Twitter platform was developed to identify potentially unsafe user behavior. In this work, four different bot versions collected feedback data after stimuli directed to 1287 Twitter accounts for 38 consecutive days. The results were not conclusive about the existence of predictors for unsafe behavior, but we conclude that despite Twitter's security policies, this kind of attack is still feasible.*

## 1. Introduction

Currently, several methods of social engineering attacks are used by attackers to gain an advantage over victims. In addition, to direct contact with victims it is commonly possible to use devices such as e-mail [Oliveira et al. 2017, Jakobsson and Ratkiewicz 2006], fake website [Jakobsson et al. 2008], automated social engineering [Shafahi et al. 2016, Lauinger et al. 2010, Huber et al. 2009, Jakobsson and Ratkiewicz 2006], obtaining information by searching garbage [Mitnick and Simon 2003, p. 126], peeking over a user's shoulders [Mitnick and Simon 2003, p. 176] [Purkait 2012], and many others [Hatfield 2018, Mitnick and Simon 2003]. These attacks act on the so-called *weakest link* or the user [Schneier 2001] [Mitnick and Simon 2003, p. 3].

In social engineering, the attacker uses influence and (or) persuasion to impersonate someone else or persuade the victim to perform actions that the victim would not normally do [Mitnick and Simon 2003, p. vii]. Social engineering on the Internet can make users provide sensitive information or facilitate malware infection. Phishing is a social engineering technique that can be defined as a scalable fraud, where personification is used to gain influence (and/or persuade) to collect information from the victim that would not normally be provided [Lastdrager 2014]. Thus, the social engineering attack via phishing can be divided in two stages: (i) personification; and (ii) information collection. It is common to use bots that perform phishing attacks in an automated manner, which substantially aggravate the situation. These types of attacks represent a serious threat to information security, as they increase the range of the attacks [Lauinger et al. 2010] making them cheaper [Huber et al. 2009].

Nowadays, artificial intelligence is advanced enough to produce bots that are capable of posing as human beings [Shafahi et al. 2016]. This type of bot is called a social bot. According to Rouse (2013), a social bot is a software that is capable of simulating human behavior through automated interactions on an online social network. Bots can be used on social networks with good motivations (e.g., a chatbot that responds respectfully to interactions with customers from some organization), but can also be used for doing malevolent tasks such as sharing spam, vectors for phishing or fake news dissemination [Freitas et al. 2015].

As related work we may cite Shafahi et al. (2016) that report the use of bots to simulate human behavior for phishing tests on Twitter. In Lauinger et al. (2010) attacks that simulate human behavior performed on Internet relay chat are described. In Huber et al. (2009) automated social engineering attacks on Facebook social media are described. Jakobsson and Ratkiewicz (2006) study attacks where the target is an online auction site, in addition they promote a discussion on the ethics involved in carrying out this type of scientific research. Shafahi et al. (2016) work is the most similar to ours since they use the Twitter platform, but their work focuses on measuring the risk that Twitter brings to corporate networks.

Among the various social media options, we can highlight Twitter, which is very popular worldwide. In January 2019, Twitter contained about 326 million active user accounts per month. In 2016, Brazil was one of the countries with the highest growth in Twitter accounts [Oliveira 2017]. One of the threats to Twitter users is phishing, an already known problem, but there is difficulty in detecting this type of attack [Purkait 2012, Jakobsson et al. 2008, Hatfield 2018]. There are already clear demonstrations that current anti-phishing tools do not work efficiently [Jakobsson et al. 2008, Purkait 2012]. However, some studies have already shown that the work that aims to educate people against this threat has good results [Alencar et al. 2013, Purkait 2012].

Therefore, this work's premise assumes that to obtain more impactful results against phishing, it is necessary to know more about this form of attack. The main problem addressed is to identify account data that indicate which are the most vulnerable to a phishing attack on Twitter. In order to identify such data, bots were created to collect information. Thus, the main contribution of this work is the description of the design process, including a bot architecture that simulates automated social engineering attacks on Twitter. The designed bot has a great chance of acting continuously suggesting that the Twitter platform is not fully prepared to mitigate such abuses.

The rest of the article includes in Section 2 the adopted methodology; in Section 3, we describe four experiments with their results; in Section 4, we present a short discussion, followed by conclusions in Section 5.

## 2. Methodology

The research was carried out within strict scientific and ethical criteria, observing the difficulty of the phenomenon to be investigated, aiming to test concepts about the operation of social engineering and phishing. The attacks were simulated, i.e. they were carried out in the most innocuous way possible, without collecting any data eventually reported by the attacked users or account data that would allow their subsequent identification.

In the experiments, the participants were not explicitly warned that a scientific experiment was undergoing, aiming to prevent them from exhibiting different behavior from that which would occur in their daily lives. Despite that, participants were informed implicitly about the experiment as explained in Section 4.2. The sequence of four experiments were performed with incremental degree of complexity, where the data and results obtained in an experiment indicated the features to be inserted in the following one.

### 2.1 Attack planning

The attack planning consiste on defining what would be executed to fulfill the objective of the experiment based on six steps:

1. **Tweet flow identification** - the accounts to be attacked would be chosen from the activity of publishing tweets, so that an attempt was made to identify a relevant and pertinent tweet flow (obtained through the use of keywords) to separate the possible victims, according to their interest groups;
2. **Obtaining tweet flows** - a continuous flow of tweets would be obtained on Twitter associated with interest groups, which would be stored in a buffer and be asynchronously consumed;
3. **Account sampling** - as the tweet flow was consumed, account sampling was performed identifying those that would be attacked;
4. **Personification** - by sending stimuli for the accounts to be attacked in the form of new tweets personified by the bots, offering information of potential interest to the users of these accounts. Anonymous identifiers were generated, and some account attributes were stored and linked to these identifiers (a theme of the tweet, number of followers, number of followers, number of posts, length of existence, and location information);
5. **Analysis of the response to stimuli** - for users who "took the bait" from the tweets, they would be presented with a page that supposedly sought to complete the attack, through the collection of personal data; and
6. **Accounting** - user behavior report (e.g., visit to the page after the attack, offer of personal data, visit the website page) generation.

## 2.2 Account samplings

Sampling and identification of accounts were based on the FollowerRank, which is an indicator used to measure the relevance of the article's author. The use of FollowerRank had its applicability demonstrated in the work of Nagmoti et al. (2010). Equation 1 defines how to calculate FollowerRank, where $i(a)$ is the number of followers of an account, and $o(a)$ the number of accounts that an account follows.

$$FR(a) = i(a) / i(a) + o(a) \qquad (1)$$

The initial hypothesis was that there could be some correlation between an account's FollowerRank and its user's vulnerability to social engineering attacks. For each tweet captured, the FollowerRank of the corresponding account was calculated, and these were segmented into FollowerRank bands. Each account in a range would receive a different attack bait, accessible through a tweet containing a URL pointing to a page that should be visited by the account user. An attribute in the URL allowed linking to the anonymized account identifier.

## 3. Experiments and results

Each experiment demanded knowledge of how Twitter works, and which strategies would be sufficient to make the bots work.

## 3.1 Experiment 1

One of the main objectives of Experiment 1 was to identify the thematic areas that would be explored in the identification of victims, based on journalistic texts that condense the annual reports on the use of the Twitter platform in Brazil. The most-commented topics were distributed among musical events, reality shows, major sporting events, and investigations by the Federal Police [de São Paulo 2017, Alves 2017]. The most popular subjects belonged to three areas: entertainment, sports, and politics. These were the theme areas chosen for the other experiments.

After identifying the areas, keywords were screened to identify possible victims for the experiments. There were 30 keywords in each thematic area. Experiment 1 performed a collection of 32.6 GB of tweets in JSON format, referring to the chosen thematic areas. Whenever the use of a keyword in a tweet was identified, all information of this tweet was stored. This data allowed to profile the FollowerRank of the active accounts in the thematic areas.

### 3.2 Experiment 2

The second experiment tested the first phishing attack strategy. The web server developed in this experiment was the same used in all subsequent ones. This one presented a page containing a form that requested personal data of the user, for a supposed registration, aiming to present him a piece of news. All accesses were counted.

The page had two buttons: register and access; and access without registration. There was also a text: "To see the project of this scientific research click here". This last URL directed the user to read the real research project reported here. The eventual access of users to the research project document was accounted. Figure 1 (in Portuguese) presents an example of the page presented to users.

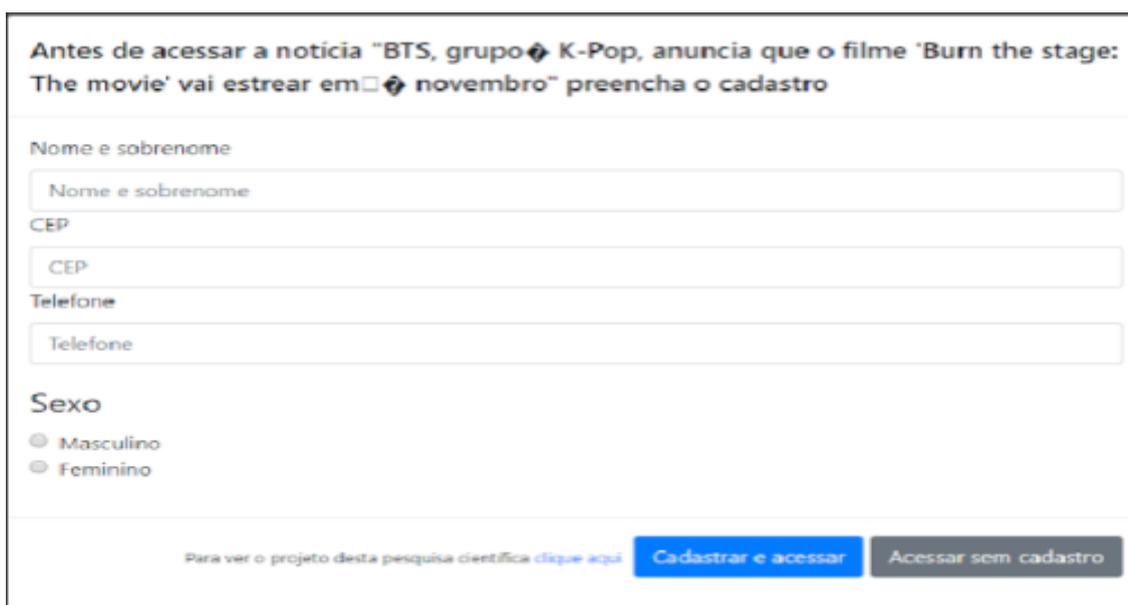

**Figure 1. Example page shown to user responding to attack.**

When clicking on the button "Access without registration", regardless of filling in the form data, the user was redirected to a legitimate page, which contained real information, addressing a related subject to one of the thematic areas. The possible access of the user to the news was counted, without the registration. The "Register and access" button was enabled only when non-null data were filled in all fields of the form. As with the "Access without registration" button, the "victim" in this case was directed to the same legitimate page. Possible access to news after the supposed realization of the register was also accounted for. The data eventually informed by user in the form was not sent to the server, nor was it stored in any way.

The success of a phishing attack depends on a personification attractive enough that the victims are tempted to act inappropriately. The personification was based on the use of real news, obtained directly and automatically, from a very popular news portal in Brazil. When implementing Experiment 2, the bot always used the first news item on the editorial page, referring to the topic of the potential interest of a user, given that his Twitter account had previously tweeted about that theme.

Experiment 2 was carried out on two batteries, one short, 30 minutes, on the eighteenth day of the research, and the other long, with a duration of three hours, on the nineteenth day of the research, when the accounts that were being used by the bots were banned by the Twitter platform.

The attacks lasted only a few hours. Two different accounts were used, where each account made specific posts in one of the thematic areas: sports or entertainment. 65 tweets (32 in one account and 33 in another) were sent to different users. The researchers chose not to carry out tests on the subject of politics, as Experiment 2 was carried out during the period of the Brazilian election campaign of 2018. During three days of execution of the pseudo-attacks of Experiment 2, the 65 attacks sent stimulated 51 hits. Most of these accesses came from news about sports, totaling 50 accesses, while only one access was made through news related to entertainment.

After performing Experiment 2, it was understood that the bots did not meet the established functional requirements. It should be much more subtle to pass unnoticed by Twitter's detection mechanisms. This fostered a refinement in the bots' requirements. Through a new reading of the use rules of the Twitter API, it was noticed that we were failing in the following points: (i) all the tweets mentioned the attacked accounts; (ii) many posts were duplicated or very similar; and (iii) the bot redirected users to an intermediate page before sending to the news.

### 3.3 Experiment 3

The purpose of Experiment 3 was to solve the flaws of the previous experiment. For each of the three failures of Experiment 2, a mitigation technique was adopted. All bot tweets mentioned another account, whose mitigation consisted of posting other tweets, interspersed with pseudo-attack tweets. Many posts were duplicated or similar, which motivated the diversification of the baits sent, combined with the posting of legit tweets (not attacks). The bot performed the redirection to an intermediate page. As it was not possible to avoid redirection we used an URL shortener.

The attacks occurred for two uninterrupted days. The tests were suspended due to another blocking of one of the accounts used, among the three different accounts used by three bots, each with specific posts in one of the thematic areas: politics, sports, or entertainment. It was unclear why the account was banned. Three possible reasons were raised to explain this ban: a target made a complaint; the bot worked 24 hours a day, without pause; and the bot used to post frequently.

Experiment 3 did not produce user interaction data that were relevant. However, there were strong indications that the development of the experiments were progressing satisfactorily. Despite the ban on one of the accounts, that account was much longer running, compared to Experiment 2. Table 1 presents the number of attacks sent in Experiment 3, segmented by thematic area.

| Experiment | Politics | Sports | Entertainment | TOTAL |
|---|---|---|---|---|
| 3 | 336 | 160 | 245 | **741** |
| 4 | 353 | 216 | - | **569** |

**Table 1. Tweets per subject for Experiments 3 and 4.**

### 3.4 Experiment 4

Similar to what happened in Experiment 3, the lessons learned in the previous experiment served to support the improvements in the new version of the bots.

Mitigation techniques were adopted for the three possible reasons for banning the lost account in Experiment 3.

We realized that the experiments could not be performed uninterruptedly. Therefore, during the night, the experiments started to be interrupted. Besides, it was found that bots should act less frequently becoming similar to human behavior [Almeida and Gondim 2019]. The decrease in the tweeting speed occurred through a very simple solution: disregarding a significant fraction of the potential targets.

After the adjustments, the tests were resumed, and their interruption occurred at the decision of the researchers, and there was no problem with the platform differently from Experiments 2 and 3. Two different accounts were used, where each bot-controlled an account and made specific posts in one of the thematic areas: politics or sports. During the entire execution of experiment 4,569 messages were sent with URLs, distributed among political and sports subjects. If compared to Experiment 3, 17 more URLs were sent in the political subject, and 56 more URLs in the sports subject. Table 1 presents the number of attack tweets sent in Experiment 4 segmented by thematic area. Experiment 4 served to show evidence that the developed bot applies to automated social engineering attacks. The bots worked for 10 days without detection, and some even managed to impersonate human users, obtaining followers.

### 3.5 Profiling

The 1287 stimuli sent generated 955 visits by single users to the phishing website, and 15 visits to the legitimate news page, support visit to the phishing website. A visitor downloaded the document with the research project. In order to identify factors that could lead Twitter users to respond insecurely to phishing attacks, logistic regressions were used, such as the one that [Oliveira et al. 2017] used, to "profile" vulnerable users. The various models aimed to identify whether the attributes of the attacked accounts could predict the chance of users associated with these accounts accessing the data collection page. Some significant, but low value, correlations were detected, relative to the positive value for the political issue, in relation to sports, while the time of existence of the account has a negative correlation.

### 4. Discussion

The bots developed in this work are less sophisticated than those described in Shafahi et al. (2016), concerning the simulation of human behavior. However, interactions with users that signal that at least some of the bots managed to pass themselves off as human users. Also, the research reported here developed more aggressive attacks, which came into direct contact with the victims, employing mentions to their accounts. But without collecting data that would allow for later identification of the victim. The research, although it did not identify which factors would be strong predictors of users' unsafe behavior, managed to produce significant results.

The bot developed in Experiment 4 proved to be stable, carrying out attacks by about 10 days without being detected. Added to the data from Experiment 3, the number of attacks carried out was over a thousand. In other words, the main contribution of this work was the conception of an exploratory methodology to build bots capable of circumventing Twitter's automatic verification mechanisms. The results suggest that the detection of bots on Twitter does not involve linguistic means (such as detecting regular expressions), but the observation of clear non-human behavior in parameters such as frequency and continuity of tweets, duration of activity and the like. This detection approach is quite simple and general and seeks not to impact user's experience.

## 5. Conclusion

An incremental iterative methodology was proposed for the development of experiments in the construction of bots that simulate social engineering attacks on Twitter. We developed a bot that managed to work for ten days without detection in the Twitter platform, attacking users in a potentially aggressive way. The hits may be used in future experiments related to the theme. The lessons learned in Experiments 2 and 3 can be used as examples of bad practices.

From the perspective of future work, we intend to extend the experiments over a longer period with the execution on other platforms. We may observe different user's attributes to map more significant predictors of unsafe behavior considering users in face of social engineering attacks.